\documentclass[twocolumn,aps,floatfix]{revtex4}
\begin{document}
\title{Comment on ``Dynamics of a Charged Particle"}
\author{Igor V. Sokolov}
\affiliation{Space Physics Research Laboratory, University of Michigan, Ann
Arbor, MI 48109 }
\begin{abstract}
The equation derived by F.Rohrlich  (Phys.Rev.E {\bf 77}, 046609 (2008))
reproduces  Eq.(76.3) from the Landau and Lifshitz book (The Classical Theory of 
Fields). The new validity condition for this equation is inapplicable 
as it is, but once fixed it 
coincides with Eqs.(75.11-12) in the cited book.
\end{abstract}

\maketitle
Dr.  Rohrlich seems to be unaware of some of important previous
work relevant to the subject of his paper \cite{bib:R}. 

{\bf 1}. I discuss the equation of the charged particle motion
derived in \cite{bib:R} as applied to a case when the external force 
is created by an electromagnetic field \cite{comment}. First, this equation reproduces 
Eq.(76.3) from the Landau and Lifshitz book \cite{bib:L}, with the derivation 
coinciding with that found in \cite{bib:P}. Consider the {\it quotations}  
from \cite{bib:P}:

{\it ``...The modified Lorentz-Dirac equation : 
$m{\bf a}={\bf f}_{\rm ext}+t_0\dot{\bf f}_{\rm ext}$ (9.8)"}. 
Compare this with Eq.(5a) from \cite{bib:R}.  
{\it ``...We obtain: $ma^\alpha=f^\alpha_{\rm ext}+t_0(\delta^\alpha_{\ \beta}+u^\alpha u_\beta)
f^\beta_{{\rm ext}\, ,\gamma} u^\gamma$ (9.9) - the relativistic version of the 
modified Lorentz-Dirac equation"}. Compare this with the ultimate Eq.(14) from 
\cite{bib:R} - the only difference is the use of $f^\beta_{{\rm ext}\, ,\gamma} u^\gamma=\dot{f}^\beta_{\rm ext}$. 

The cited derivation found in \cite{bib:P} is not presented as original work, but, as 
being derived following \cite{bib:L}. Among the problems to be solved by 
students this survey suggests the following:{\it ``{\bf Problem 16}. Show that if the 
external force is provided by an external electromagnetic field}
$F^{\alpha\beta}_{\rm ext}$, {\it then
the modified Lorentz-Dirac equation takes the form:}
$
ma^\alpha=qF^\alpha_{{\rm ext}\ \beta}u^\beta$ $+
qt_0\left[F^\alpha_{{\rm ext}\ \mu,\nu}u^\mu u^\nu +
\frac{q}m\left(\delta^\alpha_{\ \beta}+
u^\alpha u_\beta\right)F^\beta_{{\rm ext}\ \mu}F^\mu_{{\rm ext}\ \nu}u^\nu\right]."
$ 
The latter equation is present in \cite{bib:L} as Eq.(76.3). 

Paper \cite{bib:R} does not demonstrate (as solving ``Problem 16" would) that the final Eq.(14) (the same as Eq.(9.9) 
from \cite{bib:P})  is equivalent to the well-known Eq.(76.3) in \cite{bib:L}, allowing the reader to assume that the derived Eq.(14) is new.
This equivalence is evident when the  convective derivative of the 
External Lorentz Force (ELF), $f^\alpha_{\rm ext}=q F^{\alpha}_{{\rm ext} \ \mu}u^\mu$, as present in Eq.(9.9) is expanded:
\begin{equation}\label{eq:1}
\dot{f}^\alpha_{\rm ext}=u^\gamma f^\alpha_{{\rm ext}\, ,\gamma} =q\left(u^\gamma F^{\alpha}_{{\rm ext} \ \mu,\gamma}u^\mu + F^{\alpha}_{\rm ext \ \mu}a^\mu\right),
\end{equation} 
where $u^\alpha$ and $a^\alpha=\dot{u}^\alpha$ are 4-velocity and 4-acceleration, $q$ is the particle charge and $F^{\alpha\beta}_{\rm ext}$ is the field tensor. 
Now the equation to be derived (see ``Problem 16") is obtained from Eq.(9.9)
by: (1) using the anti-symmetry of the field tensor; and (2) expressing within the accuracy of the used approximation (neglecting terms $\propto \tau_0^2$) $a^\alpha$ 
in Eq.(\ref{eq:1}) in terms of the Lorentz force: $ma^\alpha=q F^{\alpha}_{{\rm ext} \ \mu}u^\mu$. Thus, no new equation of motion for a charge is derived in \cite{bib:R}. 

{\bf 2}. If, again, we consider the particular case of the ELF and employ our Eq.(\ref{eq:1}) in order to 
expand $\dot{f}^\alpha_{\rm ext}$ in the equations in \cite{bib:R}, many of them start to look unusable. 

Particularly, 
the requirement that ``... the external force 
must vary slowly enough over the size of the charge distribution'' in \cite{bib:R} 
is only applicable to the external {\it field} (not force!). Once this consideration is 
applied to {\it force}, as is done in Ineq.(1) in \cite{bib:R},  the time 
derivative of the ELF (in the 
instantaneous rest frame) should be expanded in the way analogous to how it is done in Eq.(\ref{eq:1})  
\begin{eqnarray}\label{eq:75.10}
\frac{d{\bf f}_{\rm ext}}{dt}=\lim_{|{\bf v}|\rightarrow 0}q\frac{d}{dt}\left({\bf E}+
\frac1c[{\bf v}\times{\bf B}]\right)=\nonumber\\
=q\frac{d{\bf E}}{dt}+\frac{q}c\left[\frac{d{\bf v}}{dt}\times{\bf B}\right]\approx
q\frac{d{\bf E}}{dt}+\frac{q^2}{cm}[{\bf E}\times{\bf B}],
\end{eqnarray} 
where, again,  the acceleration is  expressed within the 
accuracy of the used approximation in terms of the ELF. 
Therefore, the validity condition, 
$\tau_0 |d{\bf f}_{\rm ext}/dt|\ll |{\bf f}_{\rm ext}|$ 
(which is given in \cite{bib:R} as Ineq.(1)) actually requires that
\begin{equation}\label{eq:75.11}
\tau_0 dE/dt \ll E,\qquad
\tau_0qB/(mc)\ll 1.
\end{equation} 

The above considerations follow Ch.75 in \cite{bib:L}. Specifically, "our" 
Eq.(\ref{eq:75.10}) proves the equivalence between Eq.(5a) in \cite{bib:R} 
(in which the radiation force equals $\tau_0d{\bf f}_{\rm ext}/dt$), 
on one hand, and the radiation force as in Eq.(75.10) in \cite{bib:L} 
(expressed in terms of the right hand side of "our" Eq.(\ref{eq:75.10})), 
on the other hand.  
The validity conditions Ineqs.(\ref{eq:75.11}) are equivalent to Eq.(75.11-12) 
from \cite{bib:L}. In contrast with the ideology of \cite{bib:R} (see the quotation above), not 
only the electric field variance, but also the magnetic field magnitude should be restrained \cite{bib:L}.

{\bf Summary}. 

Approximate equations of motion and their validity conditions presented in \cite{bib:R} are 
equivalent (i.e. approximately equal within the accuracy of the used approximation) to those found in 
\cite{bib:L}. These results of \cite{bib:R} may be treated as the intermediate steps of presented in 
\cite{bib:L} derivations, which steps, although omitted in \cite{bib:L}, are available in 
more expanded presentations, such as \cite{bib:P}.

\end{document}